\documentclass{jfm}
\usepackage{graphicx}
\usepackage{epstopdf, epsfig}
\usepackage[export]{adjustbox}
\usepackage{siunitx}

\shorttitle{How surfactants influence the drop size in sprays}
\shortauthor{R. Sijs, S. Kooij and D. Bonn}

\title{How surfactants influence the drop size in sprays}

\author{R. Sijs\aff{1}
  \corresp{\email{r.j.a.sijs@uva.nl}},
  S. Kooij\aff{1}
 \and D. Bonn\aff{1}}

\affiliation{\aff{1}Van der Waals-Zeeman Institute, University of Amsterdam, Science Park 904, Amsterdam, Netherlands
}

\begin{document}

\maketitle

\begin{abstract}
Spraying is a widely used method to produce a liquid sheet that break up into droplets of a certain size distribution. When spraying simple liquids, it is known which experimental parameters determine the droplet size distribution. For many applications however, surfactants are added, producing a hitherto unknown effect on the droplet size distribution. Using two generic types of spraying nozzles, we sprayed solutions of different types of aqueous surfactants and measured the droplet size distribution of the sprays. We find that the breakup of surfactant solutions is similar to that of pure water but results in droplets that are on average smaller. The resulting droplet size distribution can be well described using the predictions for simple liquids provided that we replace the parameter of the equilibrium surface tension with the dynamic surface tension of the surfactant solution at a surface age of 20 ms, which is the characteristic time for destabilization and breakup of a liquid sheet. By rescaling them with the mean droplet size, the droplet size distributions of water and sprays with different concentrations of surfactants all collapse onto a single curve and can be well described using the compound Gamma function found previously for pure liquids. 
\end{abstract}

\begin{keywords}
Authors should not enter keywords on the manuscript, as these must be chosen by the author during the online submission process and will then be added during the typesetting process (see http://journals.cambridge.org/data/\linebreak[3]relatedlink/jfm-\linebreak[3]keywords.pdf for the full list)
\end{keywords}

\section{Introduction}
Applying a liquid by means of spraying is an ubiquitous process in many practical applications. For many such applications it is important to precisely control the size of the droplets that are sprayed. For instance, in agriculture, crops are sprayed with pesticides and the efficacy of the treatment depends on the droplet size \citep{hislop1987can}; drops should be small enough to achieve a good deposition and coverage \citep{lake1977effect}, but large enough to prevent environmental pollution due to airborne spray drift \citep{reichenberger2007mitigation, stainier2006droplet, matthews2008pesticide}. In inkjet printing, there is a relation between the size of an ink drop and the dot after spreading and drying \citep{heilmann2000effect}. In spray painting, the drop size is important for transfer efficiency and uniformity and quality of the sprayed coating \citep{hicks1995simulation}. The width of the drop size distribution is a very important parameter in the use of pharmaceutical and medicinal sprays; droplets smaller than \SI{3}{\micro \meter} are ejected from the body during exhalation while droplets larger than \SI{10}{\micro \meter} are trapped in the respiratory system \citep{babinsky2002modeling}. Droplet size control is also important for de-icing, firefighting, fuel injection, etc. \citep{lefebvre2017atomization, bayvel2019liquid}. For all these applications, a good understanding of what determines the drop size is important, allowing to control the droplet size and in this way increase the efficiency of the spraying process. 

A recent paper of \citet{kooij2018determines} describes how we can predict and therefore influence the droplet size of sprays produced by means of much-used flat fan and hollow cone nozzles. They found that for simple fluids the droplet size depends on the nozzle geometry, flow rate and surface tension of the liquid. A decrease in surface tension causes a decrease in droplet size: it becomes easier to make smaller droplets \citep{kooij2018determines, ellis2001surface}. 

In practice, the surface tension of many types of sprays is modified by the presence of surfactants such as wetting agents, which are meant to enhance droplet spreading and sticking on target surfaces, as well as penetration into leaves and barriers for active ingredient uptake. There is a wide range of observations reported in the literature as to the effect of surfactants on the process of spraying. For example, it has been reported that surfactants can change the spray dynamics \citep{stock2000physicochemical}.  When aqueous surfactant solutions are sprayed, the surface tension of the droplets is not at equilibrium, changing as surfactants migrate to the surface; this dynamic surface tension \citep{christanti2001surface} has been reported to decrease, resulting in an increasing fraction of small droplets and a corresponding decrease in mean droplet size \citep{dombrowski1954photographic, miller2000effects}. Other authors stated that, because of their relatively slow dynamics, surfactants do not influence the droplet size distribution \citep{kooij2018determines}. \citet{ellis1999adjuvants} found that the effect on droplet size are not the same for different nozzles and surfactants. Other papers stated that surfactants can increase the mean droplet size and even suppress small droplets \citep{oliveira2013potential, al2014influence}. A complete understanding of the influence of surfactants on sprays is clearly still lacking.

This paper describes how aqueous surfactant solutions affect the droplet size distribution of sprays. To be able to compare with the previous results by Kooij \etal, we used the same flat fan and conical nozzles at various pressures to spray different types of surfactants and measured the droplet size distribution using laser diffraction. 

\begin{figure}
  \centerline{\includegraphics[width=.95\textwidth]{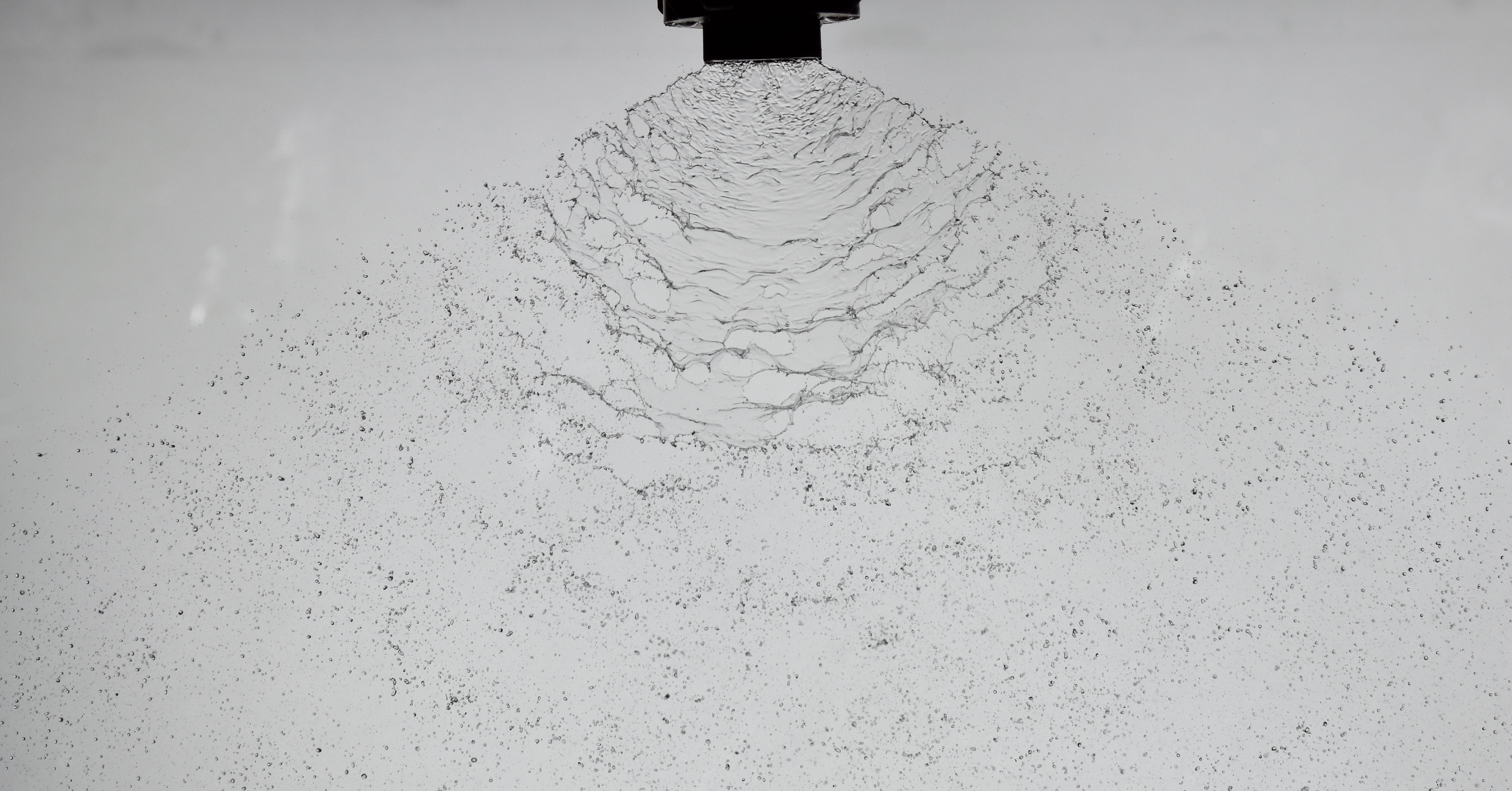}}
  \caption{Picture of water sprayed with a flat fan nozzle at an operating pressure of 3.0 bar. The sheet ruptures into ligaments which subsequently breaks up in droplets near the bottom of the picture.}
\label{fig:spray}
\end{figure}
 
\section{Droplet formation}
The formation of droplets from sprays \citep{villermaux2007fragmentation} or jets \citep{villermaux2004ligament} results from the breakup of thread-like liquid structures called ligaments. For the nozzles considered here, waves on the surface of a liquid sheet, produced by friction with the surrounding air, constitute the main breakup mechanism \citep{squire1953investigation}. These waves grow in amplitude, causing thickness modulations of the sheet to the point of rupture. This creates sheet fragments of a well-defined size, the so-called Squire wavelength, which can be observed in Fig. \ref{fig:spray}. These fragments contract to form ligaments due to the Rayleigh-Taylor instability, which is the internal instability of the sheet accelerated perpendicular to its own plane \citep{sharp1983overview}. The local diameters of the ligaments decrease until they break up into droplets; this happens because of the Rayleigh-Plateau instability that is due to the surface tension \citep{plateau1873statique, strutt1878instability}. 

The nozzles we use here operate in the regime where the Squire instability occurs, we have two relevant dimensionless parameters:

\begin{equation}
  \alpha = \rho_{air} / \rho_{liquid}
  \qquad \qquad \mbox{and\ }\qquad \qquad We = \rho_{liquid} v^2 b / \sigma ,
  \label{eq1}
\end{equation}

where, $\alpha$ is the density ratio, $We$ the Weber number, $\rho_{air}$ and $\rho_{liquid}$ are the densities of air and the liquid, respectively, $\sigma$ the surface tension, $v$ the liquid velocity and $b$ the characteristic length of the flat fan nozzle, i.e. the minor axis of the elliptical opening, derived from the hydraulic area as described by \citet{kooij2018determines}.

Using mass conservation and the fact that the median droplet size is proportional to the diameter of the ligaments, \citet{kooij2018determines} derived that the final drop size for sprays can be determined from the fluid inertia and the surface tension, or the Weber number and the geometry of the nozzle:

\begin{equation}
  D_{50} = C b \alpha^{-1/6} We^{-1/3},
  \label{eq2}
\end{equation}

where $D_{50}$ is the volume median diameter and $C$ is a constant of order unity. Equation \ref{eq2} then allows to predict the average drop size from the spraying parameters at least for pure fluids \citep{kooij2018determines}.

\section{Experimental setup}
We study the breakup of aqueous surfactant-containing sprays for a flat and conical liquid sheet. A flat fan nozzle creates a flat liquid sheet that breaks up into droplets, as can be seen in Fig. \ref{fig:spray}. A conical nozzle on the other hand, gives a sheet that is cone-shaped. Both nozzles have a round or oval inlet opening where the fluid is forced through whereafter the sheet is formed. We determined droplet size distributions using laser diffraction (Malvern Spraytec). When a laser beam hits a droplet, part of the luminous energy will be reflected, another part of the energy will be diffracted and the last part will be absorbed. The diffraction angle is inversely proportional to the size of the droplet, so the light diffraction pattern allows us, assuming a spherical shape of the droplets, to obtain the droplet size distribution \citep{swithenbank1976laser, dayal2004evaluation}. The laser beam is placed 40 cm below the nozzle, where, for all nozzles, pressures and fluid parameters, no further breakup occurs and the drops are to a very good approximately spherical. 
 
\begin{figure}
\includegraphics[width=.45\textwidth]{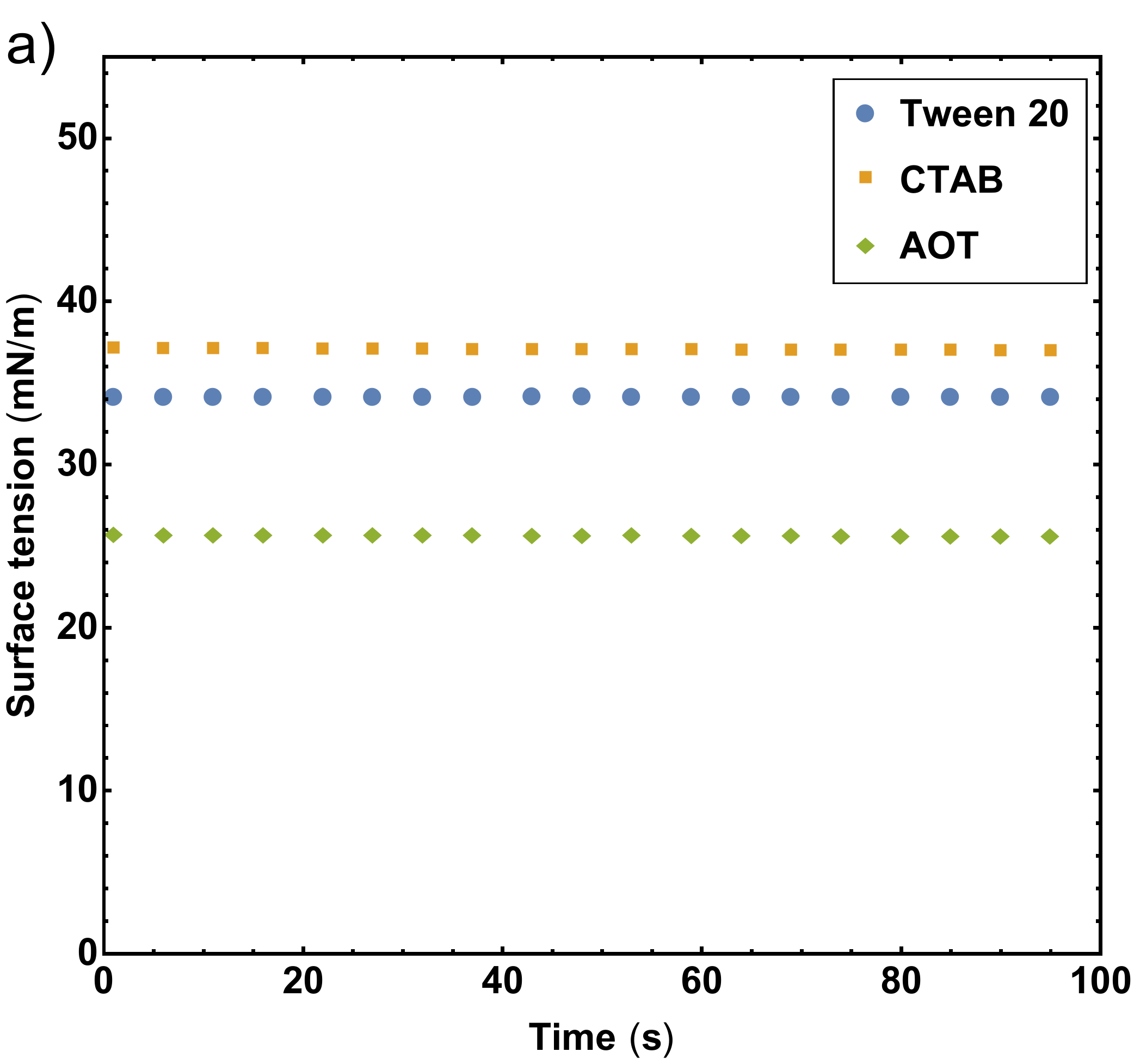} \hfill \includegraphics[width=.45\textwidth]{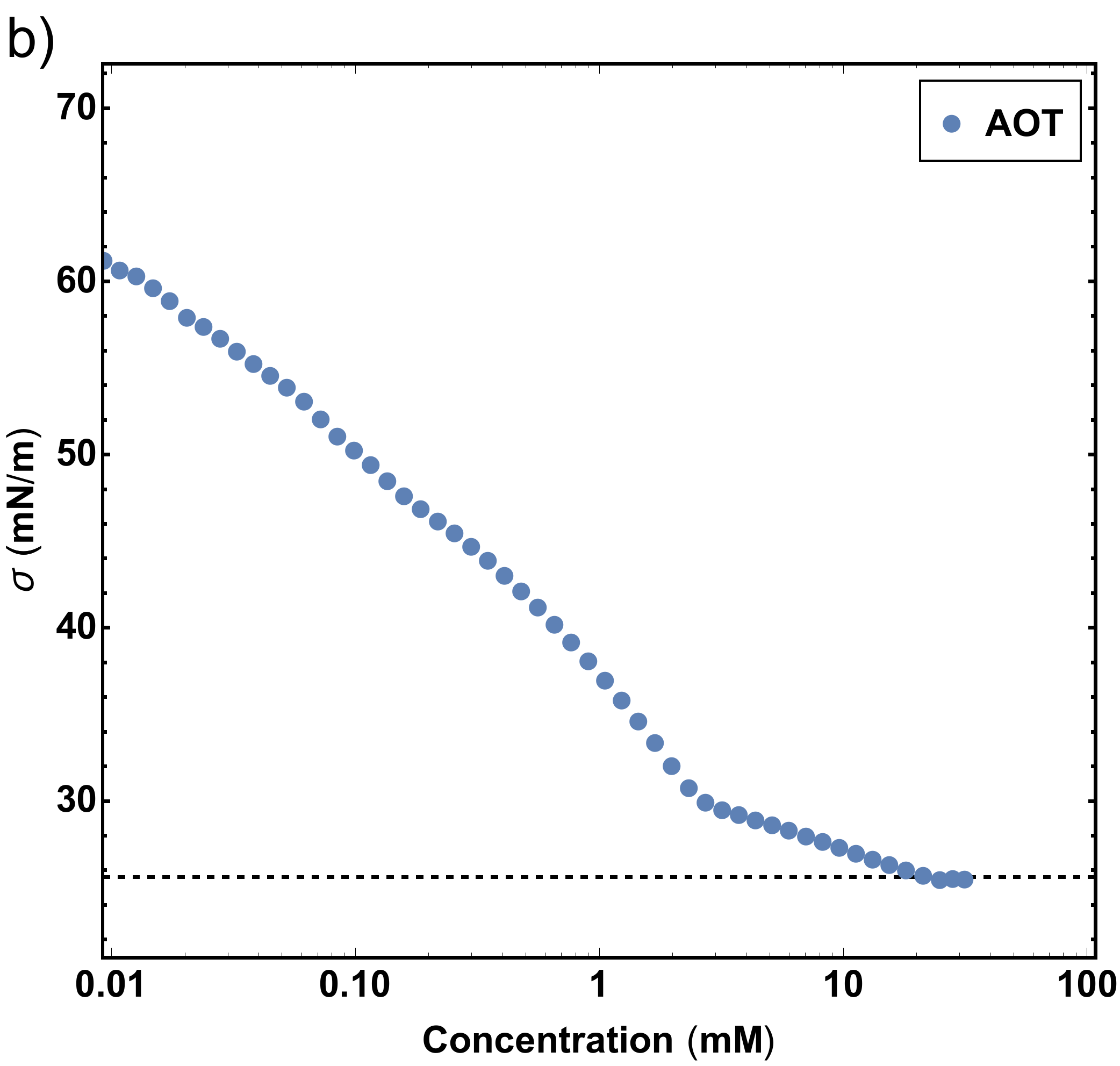}
\caption{a) Surface tension for Tween 20, CTAB and AOT as a function of time and b) the equilibrium surface tension as a function of AOT concentration.}
\label{fig:surftens}
\end{figure}

Three widely used types of surfactants are selected; a non-ionic (Tween 20), a positively charged (CTAB) and a negatively charged (AOT) surfactant, see Table \ref{tab:char}. Each has a different equilibrium surface tension ($\sigma_{equilibrium}$) as determined by a Kr{\"u}ss Force Tensiometer K100 using the Wilhelmy plate method, see Fig. \ref{fig:surftens}a, and a different critical micelle concentration (CMC), see Table \ref{tab:char} . Fig. \ref{fig:surftens}b shows the dependence of equilibrium surface tension on the surfactant concentration, in this case of AOT, which is in good agreement with earlier measurements on the same system of \citet{bergeron1996thin}. 

\begin{table}
  \begin{center}
\def~{\hphantom{0}}
  \begin{tabular}{llll}
      {\bf Surfactant}  & {\bf Surfactant type}   &   {\bf CMC (mM)} & {\bf $\bf{\sigma_{equilibrium}}$} \\[3pt]
      Tween 20 & Non-ionic & 0.05 \citep{mittal1972determination} & 34.3 mN/m \\
      CTAB & Positively charged & 1.0 \citep{neugebauer199018} & 37.0 mN/m \\
      AOT & Negatively charged & 2.5 \citep{bergeron1996thin} & 25.8 mN/m \\
  \end{tabular}
  \caption{Main characteristics of the surfactants used in the sprays}
  \label{tab:char}
  \end{center}
\end{table}

\begin{table}
  \begin{center}
\def~{\hphantom{0}}
  \begin{tabular}{lll}
      {\bf Nozzle}  & {\bf Area (m\textsuperscript{2})} & {\bf Discharge coefficient}  \\[3pt]
      Teejet XR 110-02 & 5.2 x 10\textsuperscript{-7}	& 0.94 \\
      Teejet XR 110-03	 & 8.3 x 10\textsuperscript{-7}	& 0.94 \\
      Teejet XR 110-04 & 1.1 x 10\textsuperscript{-6} & 0.91 \\
      Albuz API 110-03 & 8.8 x 10\textsuperscript{-7} & 0.85 \\
      Albuz ATR 80 (cone) & 1.1 x 10\textsuperscript{-6} & 0.34 \\
  \end{tabular}
  \caption{The used nozzles with the opening area of the nozzle and the discharge coefficient, which account for losses in the flow rate}
  \label{tab:nozzle}
  \end{center}
\end{table}

The aqueous surfactant mixtures are subsequently sprayed at pressures between 1 and 5 bar. We selected similar agricultural nozzles as used by \citet{kooij2018determines}. In Table \ref{tab:nozzle} we list their opening area and discharge coefficient, which is a parameter accounting for losses in the flow rate. The low discharge coefficient of the conical nozzle is due to the complex flow in the outlet. 

\section{Breakup process}
\begin{figure}
  \centerline{\includegraphics[width=\textwidth]{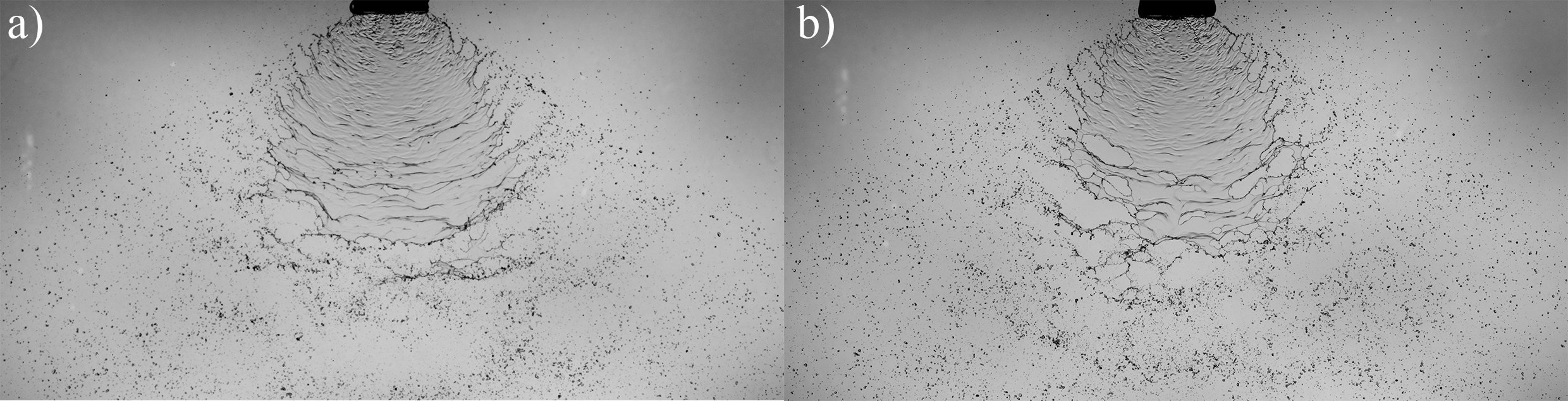}}
  \caption{High-speed photos of a) water and b) AOT at CMC in water sprays as produced by a Teejet XR 11003 VK nozzle at 2.0 bar. The flapping and breakup process are clearly visible in both cases}
\label{fig:photos}
\end{figure}

Using high-speed photography, we visualized the sheet breakup. Fig. \ref{fig:photos} shows a sheet of water (a) and water-surfactant mixture (AOT at CMC) (b), sprayed using the Teejet XR 11003 VK flat fan nozzle at 2.0 bar. It can be seen that the breakup process is qualitatively similar for both liquids. First, the sheet enters the flapping mode, which results in the formation of ligaments that subsequently break up into droplets near the bottom of the pictures.

\section{Droplet size distribution and mean drop size}
\begin{figure}
  \centerline{\includegraphics[width=.95\textwidth]{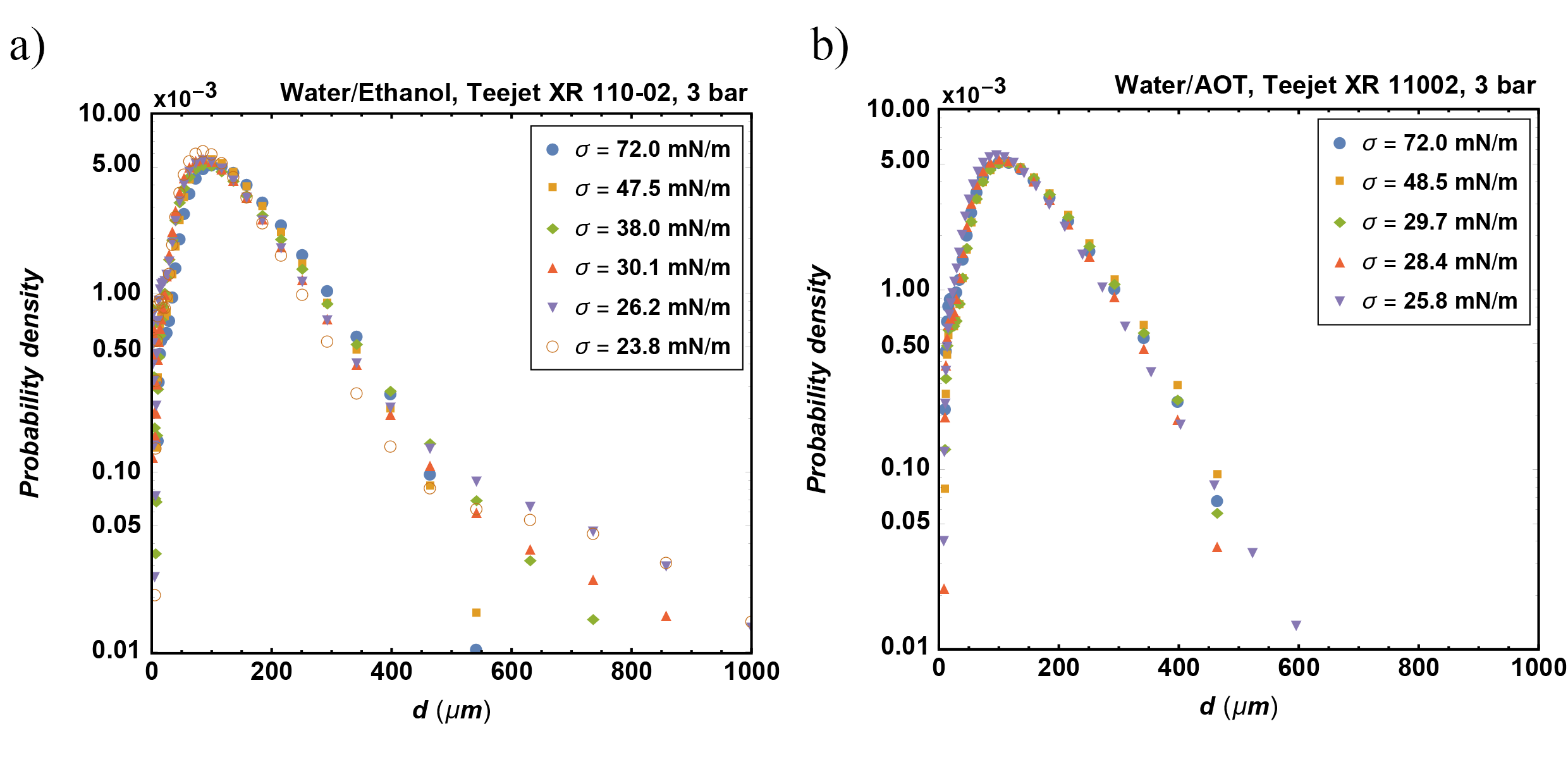}}
  \caption{The normalized droplet size distribution for a) water-ethanol and b) water-AOT mixtures. In the case of ethanol, there is a deviation for large droplet sizes. In the case of AOT, all distributions collapse onto a single curve}
\label{fig:WaterEthanolAOT}
\end{figure}

To investigate the influence of equilibrium surface tension, we sprayed different concentrations of water-ethanol and water-AOT mixtures, using the Teejet XR 11002 flat fan nozzle at an operating pressure of 3 bar. The water-ethanol mixtures allow to vary the equilibrium surface tension; for the water-AOT mixtures dynamics surface tension effects will become important too. Fig. \ref{fig:WaterEthanolAOT}a shows that for the water-ethanol mixtures, the droplet size distribution changes with surface tension; distributions differ for larger droplet sizes. For water-AOT mixtures (Fig. \ref{fig:WaterEthanolAOT}b), all distributions of different surfactant concentrations collapse and the different equilibrium surface tensions: the presence of AOT does not influence the droplet size distribution.

\begin{figure}
  \centerline{\includegraphics[width=.70\textwidth]{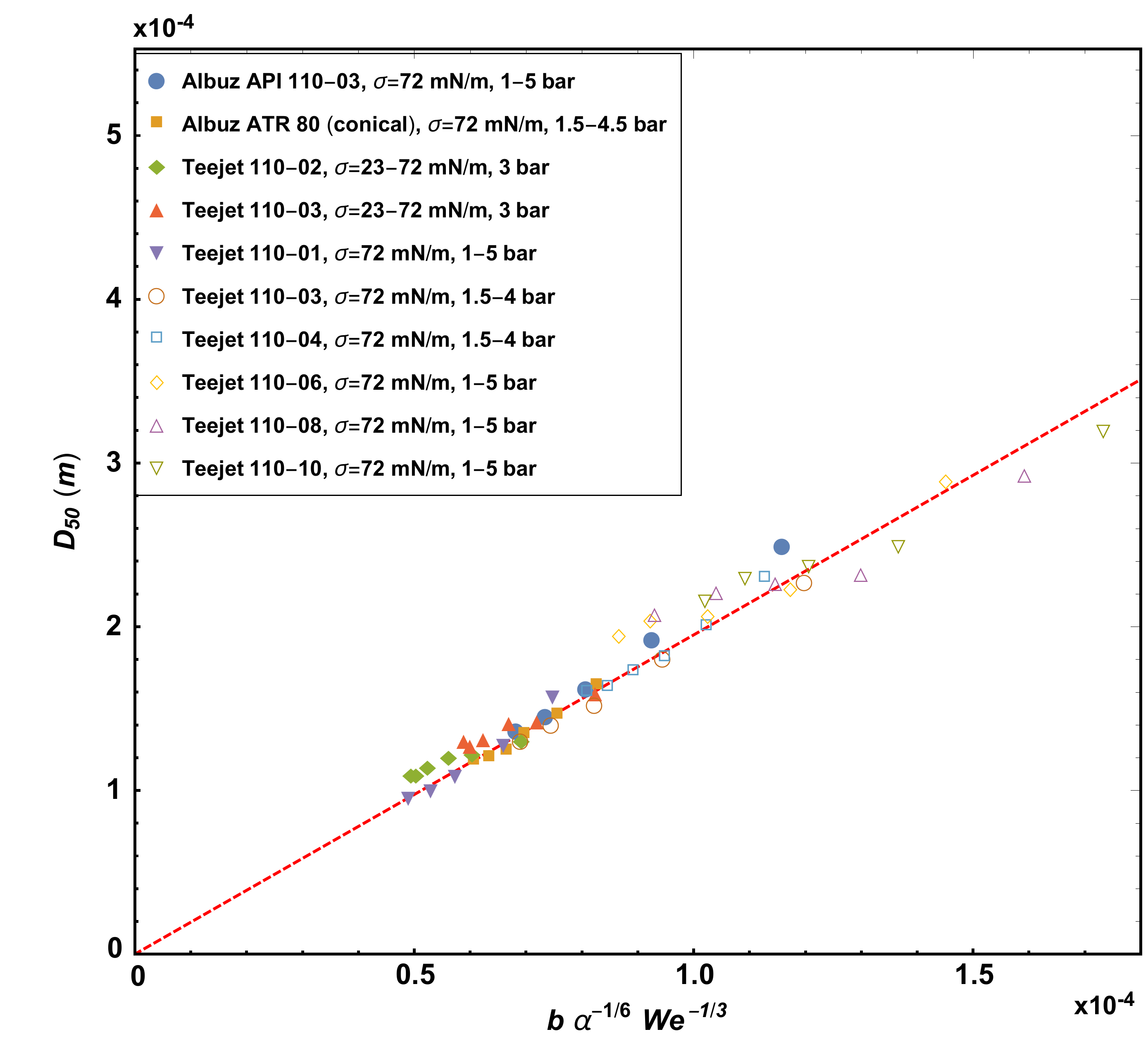}}
  \caption{The median drop size, $D_{50}$, plotted with Eq. \ref{eq2} for water and water-ethanol mixtures with different flat and conical nozzles and various pressures. The red dotted line is a fit to the data with slope $C = 1.95$}
\label{fig:D50Kooij}
\end{figure}

Determining the volume median diameter, $D_{50}$ , is the most common way to characterize the droplet size distribution in sprays. For simple water sprays and water-ethanol sprays with lower surface tension, all data have been reported to follow Eq. \ref{eq2} with $C = 1.95$ found by \citet{kooij2018determines}. Fig. \ref{fig:D50Kooij} shows a similar, but extended graph with nozzles that give a coarser or finer spray than used by Kooij \etal.

\begin{figure}
  \centerline{\includegraphics[width=0.95\textwidth]{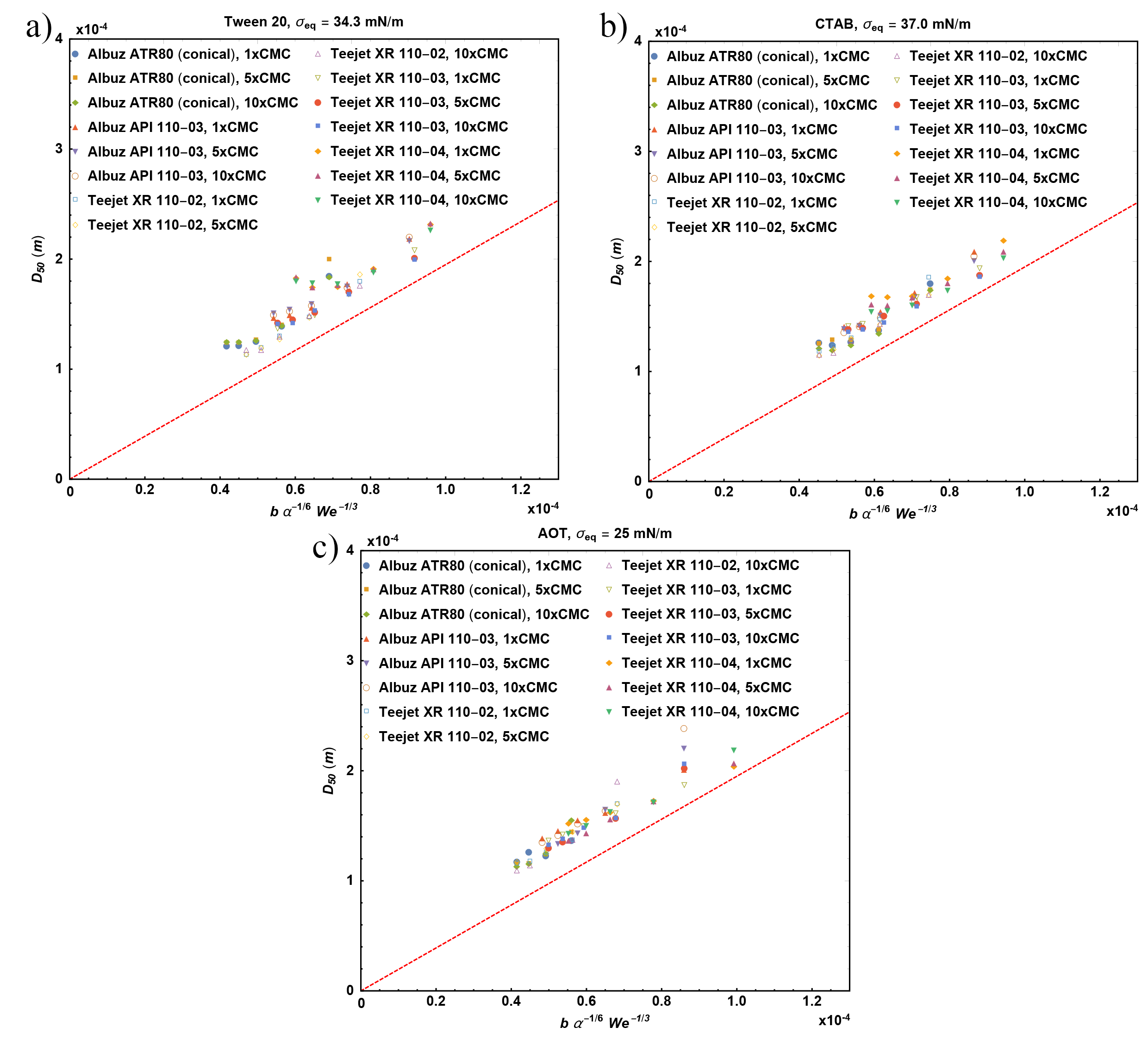}}
  \caption{The median droplet size, $D_{50}$, plotted against Eq. \ref{eq2} with the equilibrium surface tension for surfactants Tween 20 (a), CTAB (b) and AOT (c) at different concentrations, for all nozzles and various pressures between 1 to 5 bar. The dotted red line is the fit of Fig. \ref{fig:D50Kooij} with $C = 1.95$}
\label{fig:D50EST}
\end{figure}

If we plot the median drop size of the aqueous surfactant solutions against Eq. \ref{eq2} in the same way as we did for simple water and water-ethanol mixtures, using the surfactants equilibrium surface tension $\sigma$, we obtain the results shown in Fig. \ref{fig:D50EST}. All data are parallel to equation \ref{eq2} with $C = 1.95$, but are shifted to the left, and any line through the data will not go through the origin. Since $b$ and $\alpha$ are constants, this suggest that the Weber number is overestimated for all solutions. This can be explained because the equilibrium tension is used. 

\begin{figure}
  \centerline{\includegraphics[width=0.95\textwidth]{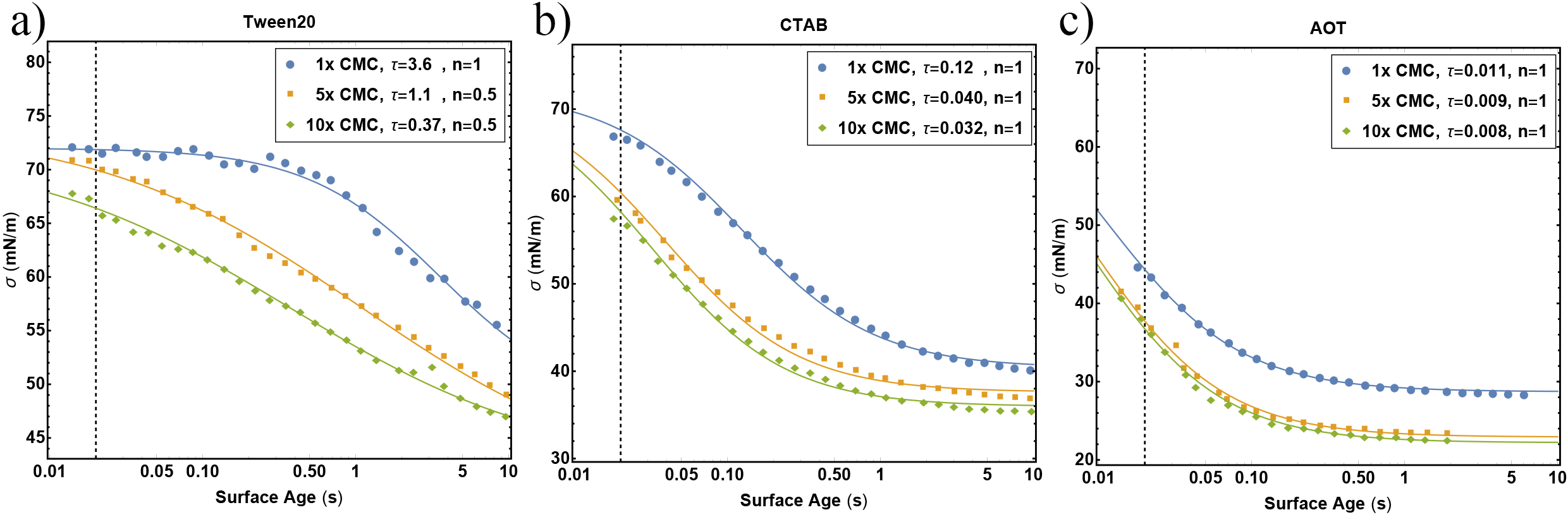}}
  \caption{The surface tension over time for various concentrations of Tween 20 (a), CTAB (b) and AOT (c). Solid lines are fits to Eq. \ref{eq3} with the parameters listed in the legends. The vertical dashed line marks 20 ms, the characteristic time corresponding to sheet breakup}
\label{fig:DST}
\end{figure}

Fig. \ref{fig:D50EST} clearly shows that the addition of surfactants results in a shift of experimental $D_{50}$ values with respect to the reference water/water-ethanol data from Fig. \ref{fig:D50Kooij}. As the Weber number is the only variable, this implies that the Weber number itself needs to decrease due to the surfactant. The density, $\rho_{liquid}$, and velocity, $v$, of the sprayed liquid are constant, as is the characteristic length $b$. The parameter left is the surface tension $\sigma$, which was taken as the equilibrium surface tension. When spraying, the fluid experiences high deformation rate flows while new surface is created rapidly during droplet breakup \citep{christanti2001surface}. Immediately after new surface is produced, the liquid will have the same surface tension as water. Then surfactant molecules move to the newly created surface area, decreasing the surface tension. When the surface is filled with molecules, the surface tension reaches the equilibrium value. 
To quantify this time-dependence, we measured the dynamic surface tension using a Kr{\"u}ss Bubble Pressure Tensiometer BP50. It provides the dynamic surface tension at surface ages of 15-16000 ms using the Young-Laplace equation. The data, shown in Fig. \ref{fig:DST}, are subsequently fitted with the equation of \citet{hua1988dynamic}:

\begin{equation}
  \sigma(t) = \sigma_\infty + \frac{\sigma_0 - \sigma_\infty}{1 + \left( \frac{t}{\tau}\right)^n},
  \label{eq3}
\end{equation}

where, $\sigma_{\infty}$ is the equilibrium surface tension, $\sigma_0$ the surface tension of water (72 mN/m), $\tau$ the characteristic time and $n \approx 1$  for surfactants \citep{aytouna2010impact}.  \citet{hewitt2002development} suggested a timescale of 20 ms as a reasonable estimate of the time of sheet breakup, meaning we should adopt the dynamic surface tension at a surface age of 20 ms in analyzing our previous results. Detailed experiments by \citet{battal2003surfactant} on surfactant adsorption on jets measured the dynamic surface tension as a function of the downstream distance; the characteristic time for a significant decrease of the surface tension due to surfactant adsorption from these measurements is $\sim$15ms, very similar to the 20 ms of \citet{hewitt2002development}.

\begin{figure}
  \centerline{\includegraphics[width=0.95\textwidth]{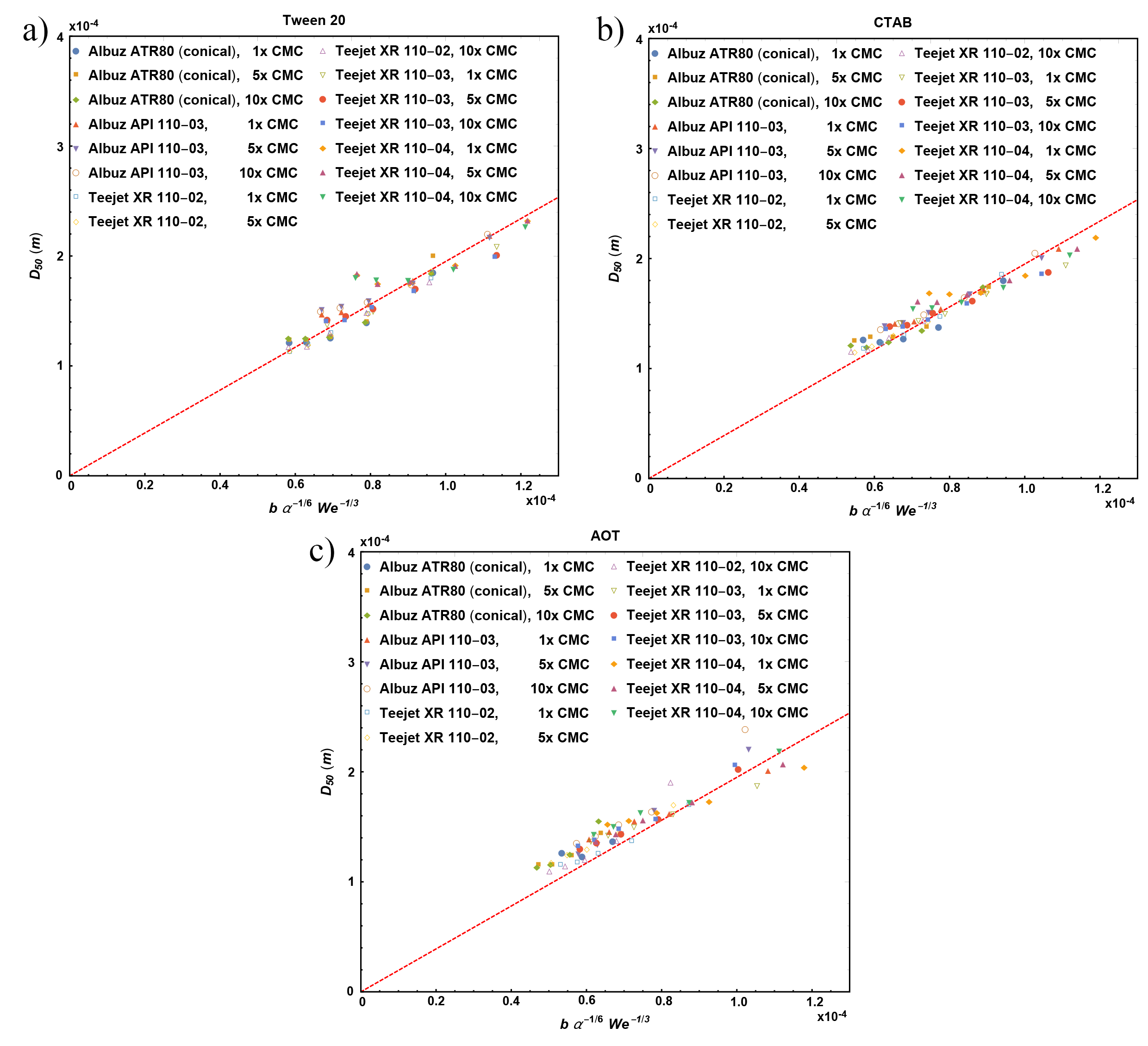}}
  \caption{The median droplet size, $D_{50}$, plotted against Eq. \ref{eq2} for surfactants Tween 20, for $\sigma_{dst, 1x CMC} = 72$ mN/m, $\sigma_{dst, 5x CMC} = 70$ mN/m  and $\sigma_{dst, 10x CMC} = 60$ mN/m  (a), CTAB , for $\sigma_{dst, 1x CMC} = 68$ mN/m, $\sigma_{dst, 5x CMC} = 60$ mN/m  and $\sigma_{dst, 10x CMC} = 58$ mN/m   (b) and AOT , for $\sigma_{dst, 1x CMC} = 44$ mN/m, $\sigma_{dst, 5x CMC} = 38$ mN/m  and $\sigma_{dst, 10x CMC} = 37$ mN/m   (c) at different concentrations, for all nozzles and various pressures between 1 to 5 bar. The Weber number was calculated using the dynamic surface tension at 20 ms. The dotted red line is the fit of Fig. \ref{fig:D50Kooij} with $C = 1.95$}
\label{fig:D50DST}
\end{figure}

Fig. \ref{fig:DST} shows that the surface tension decreases with surface aging time, meaning that the Weber number will increase and that the dynamic surface tension at 20 ms is significantly larger than the equilibrium value. Fig. \ref{fig:D50DST} shows the median drop size of the surfactant solutions when using the dynamic surface tension at 20 ms. It can be seen that now all data are well described by Eq. \ref{eq2} using the same parameter as determined for water ($C = 1.95$). 

\section{Shape of the droplet size distributions}
\begin{figure}
\includegraphics[width=.49\textwidth]{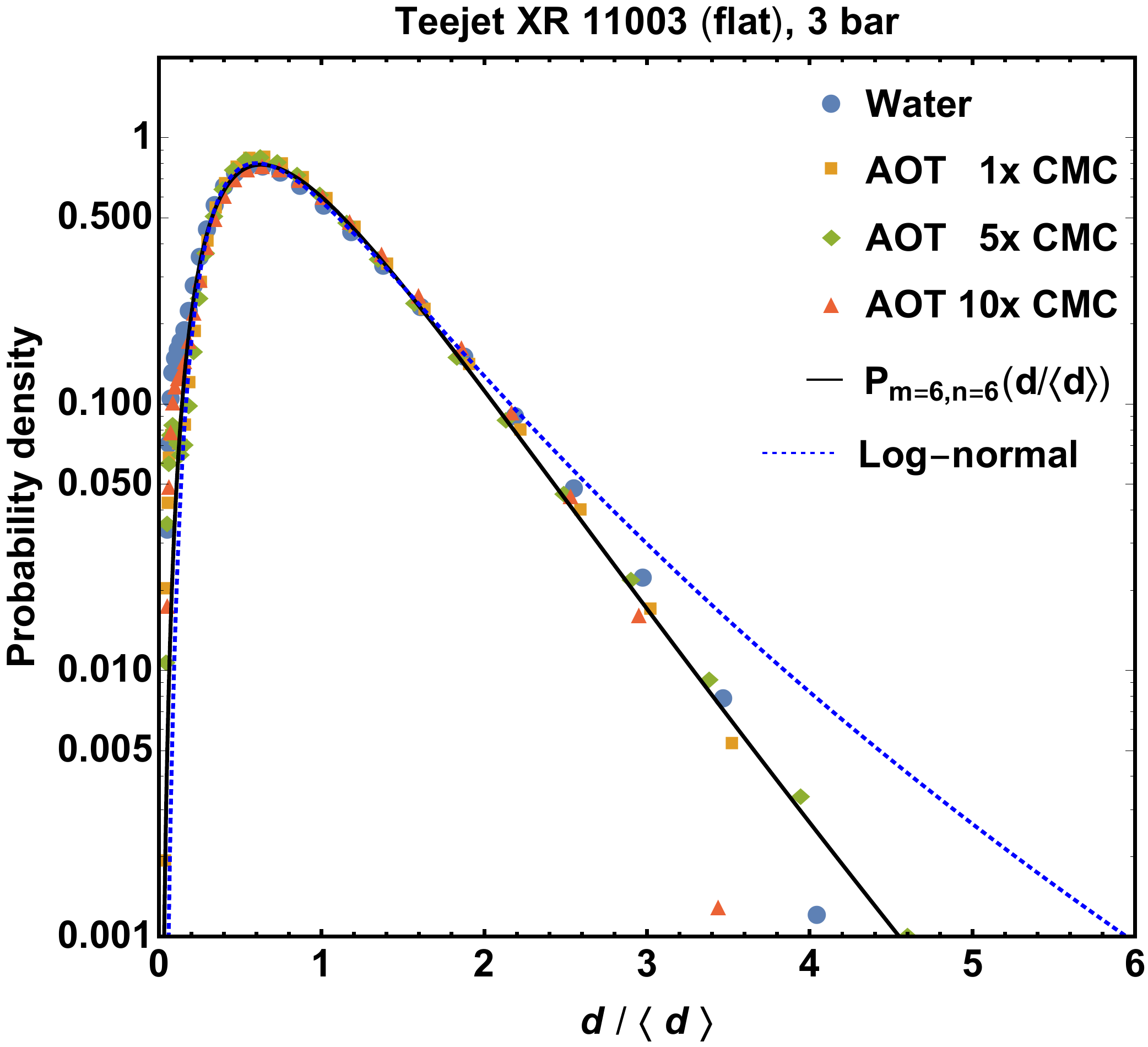} \hfill \includegraphics[width=.49\textwidth]{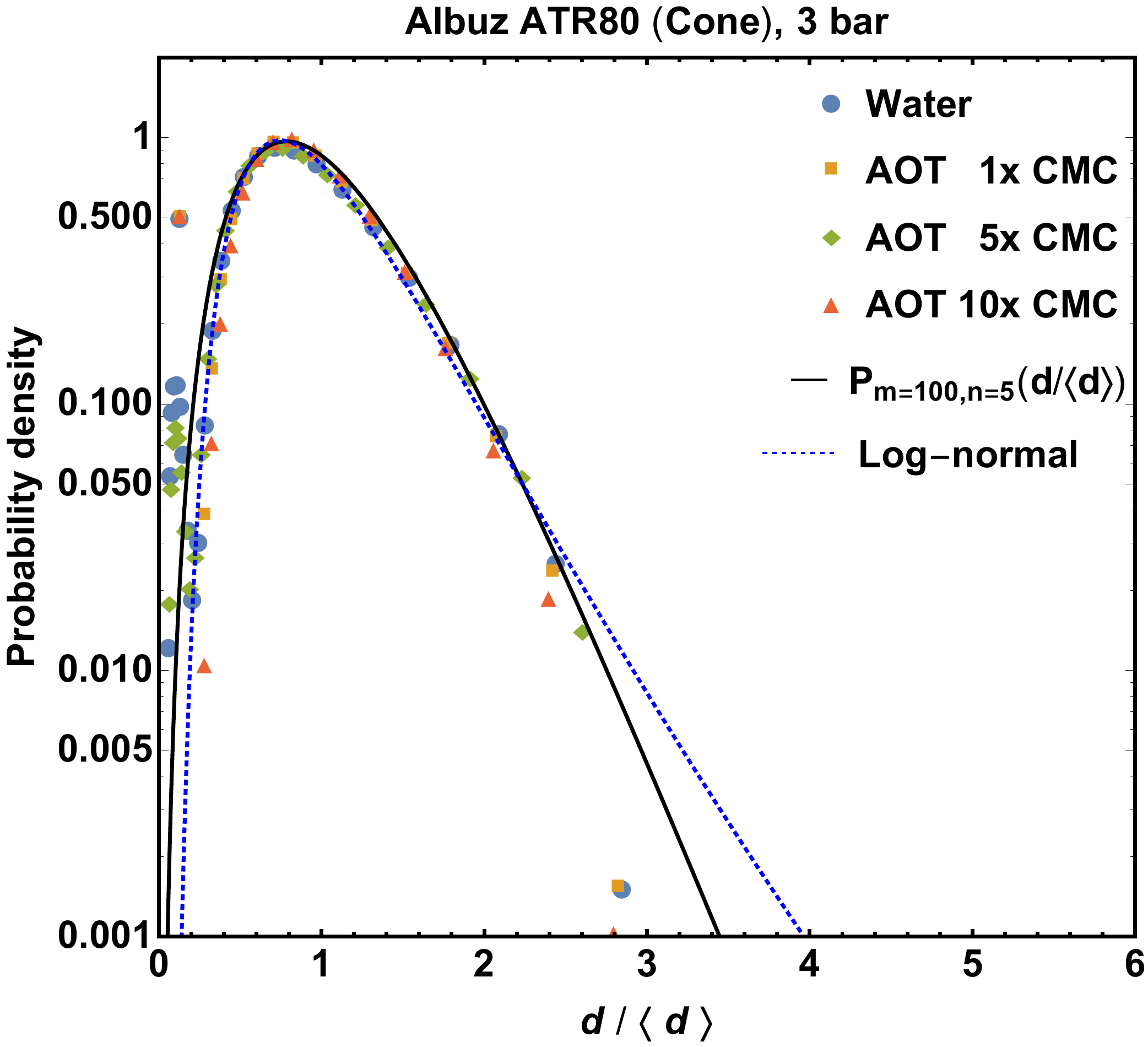}
\caption{The normalized droplet size distribution rescaled with the mean droplet size for different concentrations of AOT in water, for the flat Teejet XR 11003 (left) and conical Albuz ATR 80 (right) nozzles at a pressure of 3 bar. The data collapsing onto a single curve in both cases suggests that surfactants do not change the overall shape of the droplet size distribution. The black lines are fits using the compound Gamma function (Eq. \ref{eq4}), the blue lines to the log-normal distribution function}
\label{fig:distribution}
\end{figure}

\citet{villermaux2007fragmentation} has shown that the Gamma distribution best describes the breakup of sprays. With the breakup of a sheet, ligaments of different diameters are formed, which are also corrugated; to account for these two different parameters the distribution can best be described by the two-parameter compound Gamma function \citep{villermaux2011drop}:

\begin{equation}
  \mathcal{P}_{m,n}\left( x = \frac{d}{\langle d \rangle} \right) = \frac{2 ( mn )^{\frac{m+n}{2}}  x ^{\frac{m+n}{2}-1}}{\Gamma(m)\Gamma(n)} \mathcal{K}_{m-n}\left(2\sqrt{mnx}\right),
  \label{eq4}
\end{equation}

with $\mathcal{K}$ the modified Bessel function of the second kind and $d \,  (\langle d \rangle)$ the (mean) droplet size. The parameter $m$ sets the order of the ligament size distribution and $n$ the ligament corrugation. When rescaling the distribution by the mean droplet size, the data of different concentrations of the different surfactant collapse onto a single curve, as can be seen in Fig. \ref{fig:distribution}. The black line is the best fit of the compound Gamma function and the dotted blue line of the log-normal distribution. It can be seen that the surfactants do not significantly change the shape of the droplet size distribution. For the flat fan nozzle, the ligaments are very corrugated ($n = 6$) and of a broad range of sizes ($m = 6$). The conical nozzle produces similarly corrugated ligaments ($n = 5$) but of very similar size ($m = 100$). 

\section{Conclusion}
When spraying simple water and water-ethanol mixtures, it is possible to use the equation of \citet{kooij2018determines}, Eq. \ref{eq2}, to predict the mean droplet size of the spray based on the equilibrium surface tension. However, for sprays of aqueous solutions of surfactants, this procedure results in a shift of the data compared to the prediction. This shift can be explained by the fact that the relevant surface tension is not the equilibrium surface tension. When spraying, the fluid experiences high deformation rate flows with which new surface is created. The dynamic surface tension tells us that directly after new surface is formed the surface tension is that of water, decreasing over time until the equilibrium value has been reached. Replacing the equilibrium surface tension in Eq. \ref{eq2} with the dynamic surface tension at a characteristic time scale associated with sheet breakup estimated at $\sim$20 ms, we find that all data collapse onto a single curve following Eq. \ref{eq2} with $C=1.95$. 

By considering the droplet size distribution as a function of droplet size divided by the mean droplet size, we also show that the shape of the droplet size distribution is not changed significantly by the presence of surfactants. The rescaled droplet size distributions of water and sprays with different concentration of surfactants all collapse onto each other and can be well described using a compound Gamma function. Finally, high-speed photography shows that the breakup of surfactant sprays is qualitatively similar to that of simple water sprays. We conclude that surfactants only influence the droplet size distribution of sprays to the extent that it causes a small decrease of the mean droplet size. \\

This work is part of an Industrial Doctorate contract between the University of Amsterdam and the company GreenA B.V. that is supported by the Netherlands Organization for Scientific Research (NWO) under projectnumber NWA.ID.17.016. 

\bibliographystyle{jfm}
\bibliography{Surfactantpaper}

\end{document}